\documentstyle[11pt,epsfig]{article}
\renewcommand{\baselinestretch}{2}
\newcommand{\be}{\begin{equation}}
\newcommand{\ee}{\end{equation}}
\newcommand{\bea}{\begin{eqnarray}}
\newcommand{\eea}{\end{eqnarray}}
\newcommand{\ts}{\textstyle }

\begin{document}
\pagestyle{plain}
\def\nobreak{\penalty10000}
\flushbottom \pagenumbering{roman}

\begin{titlepage}
\renewcommand{\baselinestretch}{1.5} \small\normalsize
 \title{\bf Towards the Classification of Scalar Non-Polynomial
Evolution Equations: Polynomiality in Top Three Derivatives }
\author {{\bf  Eti M\.IZRAH\.I}\\
        {\it  Department of Mathematics, Istanbul Technical University}\\
   {\it  Istanbul, Turkey}\\
        {\it  e-mail: mizrahi1@itu.edu.tr}\\ {}
        {\bf  Ay\c{s}e H\"{u}meyra B\.ILGE}\\
        {\it  Department of Mathematics, Istanbul Technical University}\\
        {\it  Istanbul, Turkey}\\
        {\it  e-mail: bilge@itu.edu.tr}\\ {}
} \maketitle \nopagebreak[2]

\footnotetext[1]{{\bf Acknowledgment} This work is partially
supported by The Scientific and Technological Research Council of Turkey.}

\textbf{Keywords:} Evolution equations, Integrability, Classification, Symmetry, Conserved density.

\begin{abstract}\baselineskip=12pt
We prove that arbitrary (non-polynomial) scalar evolution
equations of order $m\ge 7$, that are integrable in the sense of
admitting the canonical conserved densities $\rho^{(1)},$ \
$\rho^{(2)},$ and $\rho^{(3)}$  introduced in [A.V. Mikhailov,
A.B. Shabat and V.V Sokolov ``The symmetry approach to the
classification of integrable equations" in `{\it What is
Integrability?} edited by V.E. Zakharov (Springer-Verlag, Berlin
1991)], are polynomial in the derivatives $u_{m-i}$ for $i=0,1,2$.
\textsf{We also introduce a grading
in the algebra of polynomials in  $u_k$ with $k\ge m-2$ over the ring of functions in
$x, t,u,\dots, u_{m-3}$ and show that integrable equations are scale homogeneous with respect to this grading.}
\end {abstract}
\end{titlepage}

\renewcommand{\baselinestretch}{2}
\pagenumbering{arabic}
\section*{1. INTRODUCTION}

\renewcommand{\theequation}{1.\arabic{equation}}
\setcounter{equation}{0}

 The classification problem for scalar integrable equations in one space dimension is
solved in the work of Wang and Sanders \cite{SW98} for the
polynomial and scale invariant case where it is shown that
integrable equations of order greater than or equal to seven are
symmetries of third and fifth order equations.  In subsequent
papers these results are extended to the cases where negative
powers are involved \cite{SW2000} \textsf{but their methods were
not applied  to equations without polynomiality or scaling
properties.}

The aim of the present work is the study of the classification
problem for arbitrary evolution equations.  In our work, we use
the ``formal symmetry" method, where the existence of certain
conserved densities is a necessary condition for integrability.
Our main result is that evolution equations that are integrable in
the sense above are polynomial in top three derivatives and have a
certain scale homogeneity property.

The ``formal symmetry" method introduced by Mikhailov, Shabat and
Sokolov \cite{MSS91} have been used to obtain a preliminary
classification of essentially non-linear third order equations and
quasi-linear fifth order equations. The classification of
essentially non-linear equations is considered later in the work
of Svinolupov \cite{HSS95}, where quasi-linear integrable equations
that are not linearizable  are found to be related to the
Korteweg-deVries and Krichever-Novikov  equations through
differential substitutions.

The first result towards a classification for arbitrary $m$'th
order evolution equations  is obtained in \cite{B2005} where it is
shown that scalar evolution equations $u_t=F[u]$, of order
$m=2k+1$ with $m \ge 7$, admitting a nontrivial conserved density
$\rho=Pu_n^2 +Q u_n +R$ of order $n=m+1$,  are quasi-linear. The
method of \cite{B2005} is not applicable to third order equations,
because for $m=3$, the canonical conserved density $\rho^{(1)}$ is
not of the generic form on which the quasi-linearity result is
based on. For $m=5$, although the generic form of $\rho^{(1)}$ is
valid, $k=2$ occurs as an exception and we cannot exclude the
existence of fifth order non-quasi-linear integrable equations, at
least with the present method.  For $m\ge 7$  the structure of
integrable equations seem to be much simpler and one may hope to
obtain a complete classification as in the polynomial case.

In the present paper we continue with the classification problem
using formal symmetries and we prove that evolution equations of
order $m=2k+1\ge 7$ admitting the canonical densities
$\rho^{(i)}$,  $i=1,2,3$, as given in Appendix A,  are polynomial
in the derivatives $u_m$, $u_{m-1}$ and $u_{m-2}$. The final
result presented in Corollary 4.6.2 gives the explicit form of a
candidate for an integrable  evolution equation as a polynomial in
$u_m$, $u_{m-1}$ and $u_{m-2}$, with coefficients as yet
undetermined functions of lower order derivatives. This result is
definitely the best one can obtain by the use of the canonical
densities $\rho^{(i)}$, for $i$ up to 3, as discussed  Remark 4.7
and  in the conclusion. Any further progress towards polynomiality
would require the computation of new canonical  densities. These
computations for general $m$ are extremely tedious and they are
deferred to future work. There is also an alternative and more
promising direction towards the classification: Our  candidates
for integrable equations  have a certain scale homogeneity
property with respect to a grading in the algebra of polynomials
in  $u_k$ with $k\ge m-2$ over the ring of functions in $x,
t,u,\dots, u_{m-3}$. This grading called the ``level grading" and
introduced in   \cite{tez}  proved to be an efficient tool for
practical computations and it is expected to allow the treatment
of the polynomiality in lower orders in a unified manner. However,
as both directions of approach to the problem require completely
different techniques, the present paper is limited to the
information that can be extracted from the existence of the
conserved densities $\rho^{(i)}$, for $i\le 3$.

The notation and terminology are reviewed in Section 2 and
classification results are given in Sections 3 and 4, where we
show that for sufficiently large $m$, integrable equations of
order $m$ are polynomial in the top $3$ derivatives, $u_m$,
$u_{m-1}$ and $u_{m-2}$. These  results are  obtained from the
requirement that the canonical densities $\rho^{(i)}$, $i=1,2,3$,
be conserved quantities. The expression of these canonical
densities are given in Appendix A.  In Appendix B, we present
general formulas for the $k$'th order derivatives of differential
functions for large $k$. Due to the restrictions on the validity
of these derivatives, the general formulas obtained in Section 4
are valid for $m\ge 19$. For $m< 19$, we have done explicit
computations with the symbolic programming language REDUCE and
obtained the same polynomiality results, as summarized in Section
5.

\section*{2. NOTATION AND TERMINOLOGY}
\renewcommand{\theequation}{2.\arabic{equation}}
\setcounter{equation}{0}

 Let $u=u(x,t)$. A function $\varphi $ of  $x$, $t$, $u$ and the
derivatives of $u$ up to a fixed but finite order will be called a
``differential function'' \cite{O93} and  denoted by $\varphi[u]$.
We shall assume that $\varphi$ has partial derivatives of all
orders. We shall denote indices by subscripts or superscripts in
parentheses such as in $\alpha_{(i)}$ or $\rho^{(i)}$ and reserve
subscripts without parentheses for partial derivatives. For
example if $u=u(x,t)$, then
$$u_t={\partial u \over \partial t},\quad
  u_k={\partial^k u\over\partial x^k},$$ while for $
\varphi=\varphi(x,t,u,u_1,\dots,u_n)$,
$$     \varphi_t={\partial \varphi \over \partial t},\quad
\varphi_x={\partial \varphi\over \partial x},\quad
\varphi_k={\partial \varphi\over \partial u_k}.$$ For  $\varphi$
as above,  the total derivative with respect to $x$ is denoted by
$D\varphi$ and it is given by
\be  D\varphi=\sum_{i=0}^n \varphi_i  u_{i+1}  + \varphi_x. \ee
 Higher order derivatives can be computed by applying the
binomial formula
\be D^{k}\varphi= \sum_{i=0}^n \left[\sum_{j=0}^{k-1} {k-1 \choose
j}\left( D^{j}\varphi_i\right) u_{i+k-j}\right] +
D^{k-1}\varphi_x.  \label{turev} \ee
The total derivative with respect to time denoted by $D_t$ is
given by \be D_t\varphi=\sum_{i=0}^n\varphi_i D^iF\ + \ \varphi_t.
\ee

We recall that a differential function $\rho$ is called a
conserved density, provided that there  is  a differential
function $\sigma$ such that $D_t\rho=D\sigma$. If $\rho$ is
polynomial in certain higher derivatives, in order to check the
conserved density condition, one can proceed with integration by
parts and require the vanishing of the terms that are nonlinear in
these highest derivatives.  In our derivations we shall use only
the vanishing of the coefficients of top two nonlinearities. As
shown in Proposition 3.1, these terms come from top 4 derivatives
in the expansion of $D_t\rho$. The general expression for
$D^k\varphi$ given by $(A.6d)$ is valid for $k\geq 7$ and as we
use the derivative $D^{k-2}F$ in the general formulas, we need
$k\ge9$ hence $m\ge 19$ for the validity of the general
expressions.

We shall denote generic functions $\varphi$ that depend on at most
$u_n$ by $O(u_n)$ or by $|\varphi|=n$. That is
$$\varphi=O(u_n) \quad {\rm or} \quad |\varphi|=n \quad {\rm if \ and \ only \ if }\quad
\frac{\partial{\varphi}}{\partial{ u_{n+k}}}=0 \quad {\rm
for}\quad k\ge 1.$$
If $\varphi=O(u_n)$, then $D\varphi$ is linear in $u_{n+1}$ and
$D^k\varphi$ is polynomial in $u_{n+i}$ for $i\ge 1$. In order to
distinguish polynomial functions we use the notation
$\varphi=P(u_n)$, i.e.,
$$\varphi=P(u_n)\quad {\rm if \ and \ only \ if } \quad
\varphi=O(u_n)\quad {\rm and } \quad \frac{\partial
^k{\varphi}}{\partial {u_n^k}}=0 \quad {\rm for \ some }\ k.$$
This distinction have been used in the expression of derivatives
given in Appendix B.

Note that the total derivative with respect to $x$ increases the
order by one, thus if $|\varphi|=n$ then $|D^k\varphi|=n+k$.
Furthermore, when $u_t=F$, with $|F|=m$,
 $D_t$ increases the order by $m$.

Equalities up to total derivatives with respect to $x$ will be
denoted by $\cong$, i.e.,
$$ \varphi\cong \psi \ \ \ {\rm if \ and \ only \ if } \ \ \
\varphi=\psi+D\eta $$

We shall repeatedly use integration by parts of the following type
of expressions.

Let $p_1<p_2<\dots<p_l<s-1$ and $|\varphi|=k<p_1$. Then \bea
\varphi u_{p_1}^{a_1}\dots u_{p_l}^{a_l}u_s&\cong& -D\left(\varphi
u_{p_1}^{a_1}\dots u_{p_l}^{a_l}\right) u_{s-1}, \nonumber \\
\varphi u_{p_1}^{a_1}\dots u_{p_l}^{a_l}u_{s-1}^pu_s&\cong&
-\textstyle\frac{1}{p+1} D\left(\varphi u_{p_1}^{a_1}\dots
u_{p_l}^{a_l}\right)  u_{s-1}^{p+1} \label{not1}.\eea These
integrations by parts are repeated until one encounters a monomial
which is nonlinear in the highest derivative.

The order of a differential monomial is not invariant under
integration by parts, but one can easily compute the order of the
nonlinear term that will result after integrations by parts.

\section*{3. GENERAL RESULTS ON CLASSIFICATION}

\renewcommand{\theequation}{3.\arabic{equation}}
\setcounter{equation}{0}

We start by a brief description of the formal symmetry method
\cite{MSS91}. If $R$ is a recursion operator for the evolution
equation $u_t=F[u]$, then any fractional power of $R$ is also a
recursion operator.  Thus starting from a recursion operator of
order 1, expanded in a formal  series in inverse powers of $D$,
one can compute the operators  $R^k$ of orders $k$, for
$k=2,3,\dots$.  Each of these operators satisfy the operator
equation
$$R^k_t+[R^k,F_*]=0,$$
where $F_*$ is the Frechet derivative of $F$, defined by
$F_*=\sum_{i=0}^m\ F_iD^i$.  It is known that for any two formal
series $A$ and $B$, the coefficient of $D^{-1}$ is a total
derivative \cite{A79}, hence the coefficient of $D^{-1}$ in each
$R^k$, denoted by $\rho^{(k)}$, is a conserved quantity.
In addition to these, for any $m$, it is known that
$$\rho^{(-1)}=F_m^{-1/m} \quad {\rm and} \quad  \rho^{(0)}=F_{m-1}/F_m$$
are conserved densities. We recall that as stated in Section
2, subscripts denote differentiation with respect to the
derivatives of $u$, that is
$F_m=\frac{\partial{F}}{\partial{u_m}}$ and $\rho_{m,n}$ denotes
$\frac{\partial^2{\rho}}{\partial{u_mu_n}}$, etc.

The explicit expressions of $\rho^{(1)}$ and $\rho^{(2)}$ for
$m\ge 5$ and of $\rho^{(3)}$ for $m\ge 7$ obtained in \cite{B2005}
are given in Appendix A.

In order to compute $D_t\rho$ up to total derivatives, we use
(2.3) and integrate by parts until we encounter a term which is
nonlinear in the highest derivative. The derivation of the
polynomiality result is based on the coefficients of top two
nonlinear terms, given by the equations $(\ref{bir})$ and
$(\ref{iki})$, in Proposition 3.3. To arrive to these expressions
we first show in Proposition 3.1 that if $F$ has order $m=2k+1$
and $\rho$ has order $n=m+l$, then, for $m\ge 19$, the top two
nonlinear terms are $u_{3k+l+1}^2$ and $u_{3k+l}^2$.

\noindent

\vskip 0.2cm \noindent \textbf{Proposition 3.1 } {\it Let
$\rho=\rho(x,t,u,\dots,u_n)$ be a differential function of order
$n$ and $u_t=$ } {\it $F(x,t,u,\dots,u_m)$ be an evolution
equation of order $m$, where $m=2k+1,$ \ $n=m+l$ and }  {\it
$k+l-1 \geq 0$.  Then, up to total derivatives $D_t\rho$ is}
\bea (-1)^{k+1}D_t\rho & \cong &
\left(D^{k+1}\rho_n-D^k\rho_{n-1}\right)D^{k+l}F-\left(D^k\rho_{n-2}-D^{k-1}\rho_{n-3}\right)D^{k+l-1}F
\nonumber\\
&&+\varphi . \label{e:prop} \eea {\it where
$\varphi=O(u_{3k+l-1})$.}

\noindent {\it Proof.} In $D_t\rho=\sum_{i=0}^n \rho_i D^i F +
\rho_t $, the highest order derivative comes from $\rho_nD^nF$,
where $\rho_n$ and $D^nF$ are of orders $2k+l+1$ and $4k+l+2$
respectively. If we integrate by parts $k+1$ times we obtain
$$\rho_nD^nF\cong (-1)^{k+1} D^{k+1}\rho_n\ D^{k+l}F,$$ where
$D^{k+1}\rho_n$ and $D^{k+l}F$ are now respectively of orders
$3k+l+2$ and $3k+l+1$. One more integration by parts gives a term
nonlinear in $u_{3k+l+1}$. Similarly one can see that in
$\rho_{n-1}D^{n-1}F$, $\rho_{n-1} $ and $D^{n-1}F$ are of orders
$2k+l+1$ and $4k+l+1$. This time, integrating by parts $k$ times,
we have
$$\rho_{n-1}D^{n-1}F \cong (-1)^{k} D^{k}\rho_n\ D^{k+l}F,$$
 where
$D^{k}\rho_{n-1}$ and $D^{k+l}F$ are both  of orders $3k+l+1$.
Thus the highest order nonlinear term in $u_{3k+l+1}$ comes from
top two derivatives in  $\rho_nD^n$ and $\rho_{n-1}D^{n-1}F$. By
similar counting arguments, one can easily see that top two
nonlinear terms are obtained from top four derivatives and the
remaining terms are of order $3k+l-1.$ \hfill $\Box$
\vskip0.2cm

\noindent  \textbf{Remark 3.2} As the general expressions for the
derivatives given in (A.6a-d) are valid for large $k$, there are
restrictions on the validity of the formula $(\ref{e:prop})$.
Since the top four terms of $D^{k+l}F$ and $D^{k+1}\rho_n$ are
needed in $(\ref{e:prop})$, from (A.6d) it follows that $k+1$ and
$k+l$ should be both larger than or equal to $7$. On the other
hand, at most top two terms of the expressions in the second
bracket in $(\ref{e:prop})$ contribute to the top nonlinearities
and it turns out that the restrictions coming from (A.6a,b) are
always satisfied and the crucial restriction is $k+1\ge7$ and
$k+l\ge 7$. Thus for $l=1$, $0$, $-1$ and $-2$, we need
respectively $k\ge6 (m\ge 13)$, $k\ge 7 (m\ge 15)$, $k\ge 8 (m\ge 17)$
and $k\ge 9 (m\ge 19)$.

We shall now give the explicit expressions of the coefficients of
top two nonlinear terms for $m\ge 19$ and $k+l\ge 7$.

 \vskip 0.2cm \noindent \textbf{Proposition 3.3} {\it Let
 $\rho=\rho(x,t,u,\dots,u_n)$ be a differential function of order $n$
 and}  {\it $u_t= F(x,t,u,\dots,u_m)$ be an evolution
equation of order $m$, where  $m=2k+1$,$n=m+l$.}  {\it Then for
$k\ge 9$ and $k+l\ge 7$, the coefficients of the top two nonlinear
terms $u_{3k+l+1}^2$ and}  {\it $u_{3k+l}^2$ in the expression of
$D_t\rho$ up to total derivatives is }
 $$(-1)^{k+1} D_t\rho\cong  \Lambda_1 \ u_{3k+l+1}^2
 +\Lambda_0\ u_{3k+l}^2+\varphi$$
 {\it where $\varphi=O(u_{3k+l-1})$ and $\Lambda_1$ and
$\Lambda_0$ given below.}

\bea  \Lambda_1&=&(k+{\ts \frac{1}{2}})
F_mD\rho_{n,n}-(k+l+{\ts\frac{1}{2}})DF_m\rho_{n,n}-F_{m-1}\rho_{n,n},
\label{bir} \\
\nonumber \\
 \Lambda_1 &=& \rho_{n,n}\   D^3F_{m}\       \left[{\ts\frac{1}{12}}\ (2k^3+6k^2l+6kl^2+2l^3+3k^2+3l^2+6kl+k+l)\right] \nonumber \\
&+&   \rho_{n,n}\   D^2F_{m-1}\     \left[{\ts\frac{1}{2}} \ (k^2+2kl+2k+2l+l^2+1)\right] \nonumber \\
&+&   \rho_{n,n}\   DF_{m-2}\       \left[{\ts\frac{1}{2}}\ (3+2k+2l)\right] \nonumber \\
&+&   \rho_{n,n}\   F_{m-3}\                    \nonumber \\
&+&  D\rho_{n,n}\   D^2F_m\         \left[{\ts\frac{1}{4}}\ (-2k^3-4k^2l-2kl^2+k^2+l^2+2kl+k+l)\right] \nonumber \\
&+&  D\rho_{n,n}\   DF_{m-1}\       \left[{\ts \frac{1}{2}} \ (1+l-2k^2-2kl)\right] \nonumber \\
&+&  D\rho_{n,n}\   F_{m-2}\        \left[{\ts \frac{1}{2}}\ (1-2k)\right]\nonumber \\
&+&  D^2\rho_{n,n}\ DF_m\           \left[{\ts \frac{1}{4}} (2k^3+2k^2l-k^2-k)\right] \nonumber \\
&+&  D^2\rho_{n,n}\ F_{m-1}\        \left[{\ts \frac{1}{2}}\ k^2\right] \nonumber \\
&+&  D^3\rho_{n,n}\ F_m\             \left[{\ts\frac{1}{12}}\ (-2k^3-3k^2-k)\right] \nonumber \\
&+&  D\rho_{n,n-1}\ DF_m\            \left[{\ts \frac{1}{2}}\ (-1+2k+2l)\right] \nonumber \\
&+&  D\rho_{n,n-1}\ F_{m-1}\                     \nonumber \\
&+&  D^2\rho_{n,n-1}\   F_m \        \left[{\ts \frac{1}{2}}\ (-1-2k)\right] \nonumber \\
&+&  \rho_{n,n-2}\  DF_m \           \left[2k+2l-1\right] \nonumber \\
&+&  \rho_{n,n-2}\  F_{m-1}\         \left[2\right] \nonumber \\
&+&  D\rho_{n,n-2}\ F_m \            \left[-2k-1\right] \nonumber \\
&+&  \rho_{n-1,n-1}\    DF_m \       \left[{\ts \frac{1}{2}}\ (1-2k-2l)\right] \nonumber \\
&+&  \rho_{n-1,n-1}\    F_{m-1} \    \left[-1\right]\nonumber \\
&+&  D\rho_{n-1,n-1}\   F_m   \      \left[{\ts \frac{1}{2}}\
(1+2k)\right].  \label{iki} \eea
{\it Proof.} The proof is a straightforward computation of the
integrations indicated in Proposition $3.1$. Writing the first
four terms in  $D_t\rho$ and keeping only the terms which
contribute to the nonlinearities $u_{3k+l+1}^2$, \ $u_{3k+l}^2$,
we get
\bea  (-1)^{k+1}D_t\rho
&\cong& \rho_{n,n}\ F_mu_{3k+l+1}\ u_{3k+l+2}                     \nonumber \\
&+&     \rho_{n,n}\ \left[F_{m-1}+(k+l)DF_m\right]\ u_{3k+l}\ u_{3k+l+2}   \nonumber \\
&+&     \rho_{n,n}\ \left[F_{m-2}+(k+l)DF_{m-1}+{\ts{k+l \choose 2}}D^2F_m\right]\  u_{3k+l-1}\    u_{3k+l+2} \nonumber \\
&+&     \rho_{n,n}\ \left[F_{m-3}+(k+l)DF_{m-2}+{\ts{k+l \choose 2}}D^2F_{m-1}+{\ts{k+l \choose 3}}D^3F_m\right]\   u_{3k+l-2}\ u_{3k+l+2}\nonumber \\
&+&     (k+1)\   D\rho_{n,n}\   F_m\    u_{3k+l+1}\   u_{3k+l+1}\nonumber\\
&+&     (k+1)\   D\rho_{n,n}\   \left[F_{m-1}+(k+l)DF_m\right]\ u_{3k+l}\    u_{3k+l+1}\nonumber \\
&+&     (k+1)\  D\rho_{n,n}\    \left[F_{m-2}+(k+l)DF_{m-1}+{\ts{k+l \choose 2}}D^2F_m\right]\  u_{3k+l-1}\    u_{3k+l+1}\nonumber \\
&+&     \left[\rho_{n,n-2}+D\rho_{n,n-1}+{\ts{k+1 \choose 2}}D^2\rho_{n,n}-\rho_{n-1,n-1}\right]\   F_mu_{3k+l+1}\   u_{3k+l} \nonumber \\
&+&     \left[\rho_{n,n-2}+D\rho_{n,n-1} +{\ts{k+1 \choose2}} D^2
\rho_ {n,n}-\rho_{n-1,n-1} \right ]\nonumber \\
&&\quad\quad\quad \times\left[F_{m-1} +(k+l)DF_m \right]\ u_{3k+l}\ u_{3k+l} \nonumber \\
&+&  \left[\rho_{n,n-3}+(k+1)D\rho_{n,n-2} +
kD^2\rho_{n,n-1}+{\ts{k+1 \choose
3}}D^3\rho_{n,n}-\rho_{n-1,n-2}-kD\rho_{n-1,n-1}\right] \nonumber
\\
&&\quad\quad\quad \times F_m\ u_{3k+l+1}\ u_{3k+l-1} \nonumber \\
&-& \rho_{n-2,n}\ F_m\ u_{3k+l}\ u_{3k+l+1}\nonumber\\
&-&\rho_{n-2,n}\ \left[F_{m-1}+(k+l-1)DF_m\right]\ u_{3k+l-1}\ u_{3k+l+1} \nonumber \\
&-& \left[\rho_{n-2,n-1}+kD\rho_{n-2,n}-\rho_{n-3,n}\right]\ F_m\
u_{3k+l}\ u_{3k+l}. \label{yildiz} \eea After integrations by
parts we get $(\ref{bir})$ as the coefficient of the first
nonlinear term $u_{3k+l+1}^2$ and $(\ref{iki})$ as the coefficient
of the second nonlinear term $u_{3k+l}^2.$ \hfill $\Box$
\vskip 0.2cm Actually the expression of $\Lambda_1$ is valid for
$m\ge 13$  and $k+l\ge 7$. It is clear that if $\rho$ is a
conserved density for the evolution equation $u_t=F$, then
necessarily one should have
$$\Lambda_1=0,\quad \quad \Lambda_0=0.$$
These two equations will be used repeatedly in order to derive a
number of necessary condition for integrability.

From equation $(\ref{bir})$ we can  easily get a
number of results pertaining the form of the conserved densities.
 In particular we can see that higher order conserved densities should be quadratic in
the highest derivative and top coefficients of the conserved
densities at every order are proportional to each other
\cite{B2005}.
\vskip 0.2cm \noindent  \textbf{Corollary 3.4} {\it Let
$\rho=\rho(x,t,u,\dots,u_n)$ be a differential function of order
$n$ \ and}  {\it $u_t=F(x,t,u,\dots,u_m)$, be an evolution
equation of order $m$, \ $m \geq 7$ and $n>m$. Then}
\be \rho_{n,n,n}=0 \ee
{\it Proof.} It can be seen that $(\ref{bir})$ uses only top two
terms and for $l>0$ it is  valid for $k+1\ge 3$.  Writing it in
the  form \be
(k+{\ts\frac{1}{2}})\frac{D\rho_{n,n}}{\rho_{n,n}}-(k+l+{\ts\frac{1}{2}})
\frac{DF_m}{F_m}=\frac{F_{m-1}}{F_m},\label{ucalti}\ee we can see
that for $n>m$ the highest order term is  $D\rho_{n,n}$ and it
follows that $\rho_{n,n,n}=0$. \hfill $\Box$
\vskip 0.3cm

\noindent \textbf{Remark 3.5} From $(\ref{ucalti})$ one can easily
see that if $\rho$ and $\eta$ are both conserved densities of
order $n$, with $\rho_{n,n}=P$ and $\eta_{n,n}=Q$, then
$\frac{DP}{P}=\frac{DQ}{Q}$, hence the ratio of the top
coefficients is independent of $x$.  If $\rho $ and $\eta$ are
conserved densities of consecutive orders say, $|\rho|=n$ and
$|\eta|=n+1$ with
 $\rho_{n,n}=P$ and
$\eta_{n+1,n+1}=Q$, then
$$(k+{\ts\frac{1}{2}})\left(\frac{DQ}{Q}-\frac{DP}{P}\right)=\frac{DF_m}{F_m},$$
hence $Q=F_m^{2/m}P$. \vskip 0.2cm

\noindent \textbf{Remark 3.6} If the partial derivatives of $F$
and $\rho$ in $(\ref{bir})$  and $(\ref{iki})$ depend at most on
$u_j$, then these equations are polynomial in $u_{j+i}$, $i>0$. In
all the subsequent computations we have used only the coefficient
of the top order derivatives.

\section*{4. POLYNOMIALITY RESULTS FOR THE GENERAL CASE}

\renewcommand{\theequation}{4.\arabic{equation}}

In this section we shall obtain polynomiality results, applying
Proposition 3.3  either directly   to a canonical density
$\rho^{(i)},$ \  $i=1,2,3$ (step 3 and 6),   or to generic
conserved densities $\rho,$ \ $\nu$ and show that one of the
canonical densities is of that generic form (see Table 1).
\textsf{These derivations involve quite complicated and technical derivations organized in 6
steps.
 The intermediate results displaying  polynomiality in $u_m$, $u_{m-1}$ and $u_{m-2}$
are presented respectively in Corollaries 4.1.2, 4.3.2 and 4.6.2, labeled as ``Result A",
``Result B" and ``Result C".}

Although the method seems to be a straightforward application of
the equations $(\ref{bir})$ and $(\ref{iki})$ the computations are
tedious and it took a number of trial and errors to discover the
right sequence of computations presented as Steps 1 through 6. The
computations are done in part analytically, in part with the
symbolic programming language REDUCE. As the dependencies on lower
order derivatives become explicit, the size of the actual
expressions of the canonical densities grow very fast and we had
to use generic forms for these conserved densities whenever
possible. At each step, we tried to deduce from equations
$(\ref{bir})$ and $(\ref{iki})$ a homogeneous linear system of
equations, whose non-singularity leads to a polynomiality result.
It is a remarkable fact in many cases these systems become
singular for $k=2$, hence our results are not applicable to fifth
order equations.

We use generic conserved densities in steps one, two and four. The
first step is to obtain the quasi-linearity result for $m
> 5$, which follows from the fact that the coefficient matrix of a
homogeneous system is non-singular for $m > 5$. At the second and
fourth steps we have a similar structure; we show that the
coefficient of $u_m$ is independent of $u_{m-1}$ and $u_{m-2}$
respectively, by obtaining nonsingular homogeneous systems of
linear equations.

The third and sixth steps are based on relatively straightforward
computations using the canonical densities.  At the third step we
complete polynomiality in $u_{m-1}$ while at the fifth and sixth
steps we complete polynomiality in $u_{m-2}$, by using the
explicit form of the canonical densities $\rho^{(1)}$ \ and \
$\rho^{(3)}$.

\subsection*{4.1. Step 1. quasi-linearity: $F_{m,m}=0$}

\setcounter{equation}{0}

We now prove that an evolution equation admitting a quadratic
conserved density is quasi-linear. The quasi-linearity result was
already given in \cite{B2005}, but we shall repeat this
derivation, because with Proposition 3.3, the proof given below is
much shorter and neater.

\vskip 0.2cm \noindent
  \textbf{Proposition 4.1.1} {\it Let $u_t=F(x,t,u,\dots,u_m)$, with \
$m=2k+1\ge13$ and assume that it admits a conserved density of the
form \be \rho=Pu_{m+1}^2+Qu_{m+1}+R \label{step1}\ee \ where
$|P|=|Q|=|R|=m$. \ Then $PF_{m,m}=0.$}
\vskip 0.2cm\noindent
 {\it Proof.} Since the conserved density has order $m+1$, the restriction  for the applicability of (A.6d)
  is  $k+1\ge 7$ hence $m\ge 13$.  The coefficient of $u_{m+1}$ in $(\ref{bir})$
and the coefficient of $u_{m+3}$ in $(\ref{iki})$ are respectively
as follows: \be\displaystyle \left(2k+1\right)F_mP_m -
\left(2k+3\right)PF_{m,m}= 0 \nonumber \ee
\be \displaystyle (2k+1)(k^2+k+6)F_mP_m - (2k+3)(k+1)(k+2)PF_{m,m}
= 0 \nonumber \ee
 These two equations form the following homogeneous system of linear equations
\bea \displaystyle \left[%
\begin{array}{ll}
  1 & -1 \\
  (k^2+k+6)& -(k^2+3k+2) \\
\end{array}
\right] \left[%
\begin{array}{l}
  F_mP_m \\
  PF_{m,m}\\
\end{array}
\right]= \left[%
\begin{array}{l}
  0 \\
  0 \\
\end{array}%
\right] \label{yirminci} \eea
The coefficient matrix in this linear system is singular for
$k=2$.  For \ $k\neq2$ \ the homogeneous system has only the
trivial solution hence from $(\ref{yirminci})$ we conclude that
\be F_mP_m=PF_{m,m}=0. \ee \hfill $\Box$ \\
The quasi-linearity result follows from here immediately.
\vskip0.05cm
\noindent
\textbf{Result A: Polynomiality in $u_m$}
\vskip 0.05cm \noindent \textbf{Corollary 4.1.2} {\it
If the canonical density $\rho^{(1)}$ is a conserved quantity for the evolution equation
$u_t=F$,
then
$F_{m,m}=0,$ hence $$u_t=Au_m+B$$
where $A$, $B$ are functions of $x,t,u,\dots u_{m-1}$.}
\vskip 0.2cm  \noindent \textsf{{\it Proof.} $\rho^{(1)}$ in $(A.3)$ is of
the form $(\ref{step1})$ with $P=a^{-1}a_m^2$ where $a=F_m^{1/m}.$
Hence
$PF_{m,m}=0$ implies $F_{m,m}=0,$ \  and \ $u_t=Au_m+B$} \hfill $\Box$ \\

In reference \cite{B2005}, by direct computation of the conserved
density conditions, it was shown that equations of orders $7$, $9$
and $11$ were also quasi-linear. Since the existence of the
canonical densities are necessary conditions for integrability, it
follows that integrable evolution equations of order $m=2k+1$ with
$k\ge 3$ are quasi-linear.
As discussed in Proposition 3.1, the applicability of the general
formulas to the computation of the conserved density conditions
require that $k+l\ge 7$, hence the results below are valid only
for $m \geq 19$.  For lower orders we have checked the validity of
our results by direct computations and outline the results in
Section 5.

\subsection*{4.2 Step 2: Polynomiality in $u_{m-1}$, first result,
$A_{m-1}=0$}

In the second and third steps we determine the dependency of the
coefficients $A$ and $B,$ in $u_t=Au_m+B$ on $u_{m-1}$. For this
purpose we consider a generic quadratic conserved density $\rho$
of order $m$ and use the coefficients of the top two
nonlinearities in $D_t\rho$ which are respectively given by
equations $(\ref{bir})$ and $(\ref{iki})$.
\vskip 0.2cm

\noindent  \textbf{Proposition 4.2.1}\ {\it Let $u_t=Au_m+B$, \
$m\geq 19$ \ with  \ $|A|=|B|=m-1,$ and assume
 that it admits a conserved density
 \be \rho=Pu_m^2+Qu_m+R,  \label{step2}\ee
 where $|P|=|Q|=|R|=m-1.$ \ Then \be
 PA_{m-1}=0. \ee
}
\vskip 0.1cm
\noindent {\it Proof.}  We use here the equations $(\ref{bir})$
and $(\ref{iki})$ with $l=0,$ \ $F_m=A$ \ and $\rho_{n,n}=2P$. The
coefficient of $u_m$ in $(\ref{bir})$ is
\bea \left(2k+1\right)P_{m-1}A-\left(2k+3\right)A_{m-1}P=0,
\label{sonbir} \eea
\noindent while the coefficient of $u_{m+2}$ in $(\ref{iki})$ is
\be
\left[2k^3+9k^2+13k+6\right]A_{m-1}P-\left[2k^3+3k^2+13k+6\right]P_{m-1}A=0.
\label{soniki} \ee
\noindent From $(\ref{sonbir})$ and \ $(\ref{soniki})$ we get the
following linear system.
\bea \displaystyle \left[%
\begin{array}{ll}
  2k+1 & -(2k+3) \\
  2k^3+3k^2+13k+6 & -(2k^3+9k^2+13k+6) \\
\end{array}
\right] \left[%
\begin{array}{l}
  AP_{m-1} \\
  PA_{m-1}\\
\end{array}
\right]= \left[%
\begin{array}{l}
  0 \\
  0 \\
\end{array}%
\right]\eea
The coefficient matrix of the linear system is singular for $k=2,$
then $m=5$ which appear as an exception. For $k\neq 2$, the system
has a trivial solution. Hence $PA_{m-1}=AP_{m-1}=0$. \hfill $\Box$

Again by using the canonical density $\rho^{(1)}$ we shall obtain
that $A$ is independent of $u_{m-1}$

 \vskip 0.2cm \noindent \textbf{Corollary 4.2.2} {\it If the
canonical density $\rho^{(1)}$ is conserved, then} $A_{m-1}=0$.

\vskip 0.1cm

\noindent {\it Proof.} We  substitute  $u_t=Au_m+B$ in $(A.3)$ and
integrate by parts we can see that  $\rho^{(1)}$ is of the form
$(\ref{step2})$ where
\be P=\frac{a_{m-1}^2}{a}, \ \ \ \  A=a^m. \ee
Hence $PA_{m-1}=0$ implies $A_{m-1}=0$. \hfill $\Box$ \\

\subsection*{4.3 \ Step 3: Polinomiality in $u_{m-1}$, second
result, $B_{m-1,m-1,m-1}=0$}

Now we shall see that the conservation of the canonical densities
$\rho^{(1)}$ and $\rho^{(3)}$ will determine the form of $B$.
By substituting $u_t=Au_m+B$, with \ $|A|=m-2$ and $|B|=m-1$ in
the canonical densities
 $\rho^{(1)}$  and $\rho^{(3)}$ one can see that they reduce to
 $$ \rho^{(1)}=P^{(1)}u_{m-1}^2+Q^{(1)}u_{m-1}+R^{(1)} $$
 $$ \rho^{(3)}=P^{(3)}u_m^2+Q^{(3)}u_m+R^{(3)} $$
where $|Q^{(1)}|=|R^{(1)}|=m-2,$ and $|Q^{(3)}|=|R^{(3)}|=m-1$,
and
\be P^{(1)}=\frac{24}{m^2-1}a_{m-2,m-2}+a_{m-2}^2a^{-1}(m^2-1),
\ee
\bea \displaystyle
P^{(3)}&=&\frac{a}{m^3+3m^2-m-3}\left[a_{m-2}^2\left(m^3+3m^2-121m+597\right)\right.\nonumber
\\
&+& \left.
60a^{-m+1}a_{m-2}B_{m-1,m-1}\left({\ts\frac{3}{m}}-1\right)
\right.\nonumber
\\
&+& \left.
60a^{-2m+2}B_{m-1,m-1}^2\left({\ts\frac{1}{m}}-{\ts\frac{1}{m^2}}\right)\right]
\label{pepe3} \eea
\vskip 0.2cm

\noindent
 \textbf{Proposition  4.3.1} {\it
\  Let $u_t=Au_m+B$, \ $m\geq 19$ \ with  \ $|A|=m-2$ and
$|B|=m-1$. Then if the canonical densities $\rho^{(1)}$ and
$\rho^{(3)}$ are
 conserved quantities,  then} \be B_{m-1,m-1,m-1}=0. \ee

 \noindent {\it Proof.} To prove this result,  it would be
actually sufficient to use $\rho^{(3)}$ only, but we first see how
far one can go by using $\rho^{(1)}$. We first compute
$(\ref{bir})$ for $\rho=\rho^{(1)}$, hence with $l=-1,$ \ $F_m=A$
\ and \ $\rho_{n,n}=\rho_{m-1,m-1}=2P^{(1)}$ \ to get
\be
(k+{\ts\frac{1}{2}})AP^{(1)}_{m-2}u_{m-1}-(k-{\ts\frac{1}{2}})P^{(1)}A_{m-2}u_{m-1}-P^{(1)}B_{m-1}=0.
\label{cicek} \ee
Differentiating $(\ref{cicek})$ twice with respect to $u_{m-1}$ we
obtain \be 2P^{(1)}B_{m-1,m-1,m-1}=0. \label{bocek} \ee
If $P^{(1)}$ is nonzero, then $B$ is quadratic in $u_{m-1}$, but
$P^{(1)}=0$ gives a differential equation for $a$ and we cannot
exclude the possibility that $B_{m-1,m-1,m-1}\neq 0$. It is
possible to solve this differential equation, but it is easier to
use $\rho^{(3)}$.

Now we compute  $(\ref{bir})$ using $\rho=\rho^{(3)}$, hence with
$l=0,$ \ $F_m=A$ \ and \ $\rho_{n,n}=\rho_{m,m}=2P^{(3)}$ \ to get
\be
(k+{\ts\frac{1}{2}})AP^{(3)}_{m-1}u_m-(k+{\ts\frac{1}{2}})P^{(3)}A_{m-2}u_{m-1}-P^{(3)}B_{m-1}=0.
\label{bortu} \ee
This equation is linear in $u_m$ and its coefficient gives  \be
(k+{\ts\frac{1}{2}})AP^{(3)}_{m-1}=0. \ee Now since $A$ should be
nonzero, it follows that $P^{(3)}$ is independent of $u_{m-1}$.
Differentiating $P^{(3)},$ in $(\ref{pepe3}),$ with respect to
$u_{m-1}$, we obtain \bea \displaystyle
B_{m-1,m-1,m-1}\left[a^{-m+1}a_{m-2}\left({\ts\frac{3}{m}}-1\right)
+2a^{-2m+2}B_{m-1,m-1}\left({\ts\frac{1}{m}}-{\ts\frac{1}{m^2}}\right)
\right]=0.\eea
Thus we can conclude that:
\be B_{m-1,m-1,m-1}=0.\ee

\vskip0.05cm
\noindent
\textbf{Result B: Polynomiality in $u_m, u_{m-1}$}
\vskip 0.05cm \noindent \textbf{Corollary 4.3.2} {\it
If the canonical densities  $\rho^{(1)}$  and $\rho^{(3)}$ are conserved quantities
for the evolution equation
$u_t=F$,
then
\be
F=Au_m+Cu_{m-1}^2+Du_{m-1}+E.\label{sonuc2}\ee
where $A$, $C$, $D$, $E$  are functions of $x,t,u,\dots u_{m-2}$.}

\subsection*{4.4 Step 4: Polinomiality in $u_{m-2}$, first result,
$A_{m-2}=C=0$}

At this step we assume the existence of two generic conserved
densities $\rho=Pu_{m-1}^2+Qu_{m-1}+R$ and $\eta=Su_m^2+Tu_m+U$
and we use the relation $S=F_m^{2/m}P$ given in Remark 3.5 and
obtain $CP=AP_{m-2}=APa_{m-2}/a=0$.  Then we compute the explicit
form of the canonical densities $\rho^{(1)}$  and $\rho^{(3)}$ to
prove that these imply that  $A_{m-2}=C=0$. \vskip 0.2cm

\noindent \textbf{Proposition 4.4.1} \ {\it Let \
$u_t=Au_m+Cu_{m-1}^2+Du_{m-1}+E,$ \ $m\geq 19,$ $m=2k+1$ \ with
$|A|=|C|=|D|=|E|=m-2,$  \ and assume that it admits two conserved
densities \be \rho=Pu_{m-1}^2+Qu_{m-1}+R, \  \ee \be \eta
=Su_m^2+Tu_m+U, \ \ \ee where \ $|P|=|Q|=|R|=m-2,$ \ $|T|=|U|=m-1$
and $S=F_m^{2/m}P.$ Then} \be CP=AP_{m-2}=APa_{m-2}/a=0 .
\label{fourtw}\ee
\vskip 0.1cm

\noindent  {\it Proof.} If we compute \ $(\ref{bir})$ and
$(\ref{iki})$ for $l=-1$ \ $F_m=A=a^m,$
$A_{m-2}=ma^{m-1}a_{m-2}=(2k+1)a^{m-1}a_{m-2}$ \ and \
$\rho_{n,n}=\rho_{m-1,m-1}=2P$ \ we obtain the coefficient of
$u_{m-1}$ in $(\ref{bir})$ \ as
\bea \displaystyle
4CP-\left(2k+1\right)AP_{m-2}+(2k-1)(2k+1)AP\frac{a_{m-2}}{a}=0
\label{sunbir} \eea
and the coefficient of $u_{m+1}$ in $(\ref{iki})$  \ as
\bea \displaystyle & & 12k^2CP
-\left(2k^3+3k^2+13k+6\right)AP_{m-2}
\nonumber \\
 &&+\left(2k^3-3k^2+13k+6\right)(2k+1)AP\frac{a_{m-2}}{a}=0.
 \label{suniki} \eea
\noindent Then for $l=0$ \ $F_m=A=a^m$ \ and \
$\rho_{n,n}=\rho_{m,m}=2S=2a^2P$ the coefficient of $u_{m+1}$ in
$(\ref{iki})$ is obtained as
\bea &&12(k^2+2k+1)PC-(2k^3+3k^2+25k+12)AP_{m-2}+
\nonumber \\
&&+(4k^4+4k^3+23k^2-19k-6)AP\frac{a_{m-2}}{a}=0 \label{sonbes}
\eea
Equations $(\ref{sunbir}),$ \ $(\ref{suniki})$ \ and \
$(\ref{sonbes})$ form the following system:
\bea \displaystyle \left[%
\begin{array}{llll} \displaystyle
  4      & -(2k+1)         & (4k^2-1) \\
 12k^2    & -(2k^3+3k^2+13k+6) & (4k^4-4k^3+23k^2+25k+6)\\
  12(k^2+2k+1)  & -(2k^3+3k^2+25k+12) & (4k^4+4k^3+23k^2-19k-6)        \\

\end{array}
\right] \left[%
\begin{array}{l}\displaystyle
 CP \\
 AP_{m-2} \\
 AP\frac{ a_{m-2}}{a} \\
\end{array}
\right]= \left[%
\begin{array}{l}
  0 \\
  0 \\
  0 \\
\end{array}%
\right]  \nonumber \\
\label{bayrak}\eea
The coefficient matrix in this linear system is singular for
$k=-\frac{1}{2},-\frac{3}{2},2.$  Notice that for $k=2,$ \ $m=5$
is also an exception. For $k\neq 2$, the system has only the
trivial solution. Thus from $(\ref{bayrak})$ we have
$(\ref{fourtw})$

 \hfill  $\Box$

Now we shall compute the canonical densities $\rho^{(1)}$,
$\rho^{(2)}$, and $\rho^{(3)}$ and we shall see that their
conservation will give $ A_{m-2}=C=0.$

By substituting $u_t=Au_m+Cu_{m-1}^2+Du_{m-1}+E,$ \ $m\geq 19,$
$m=2k+1$ \ with \ $A=a^m,$ \  $|A|=|C|=|D|=|E|=m-2,$ in the
canonical densities
 $\rho^{(1)},$ $\rho^{(2)},$ and $\rho^{(3)}$ one can see that they reduce
to
    \bea \rho^{(1)}&=&Pu_{m-1}^2+Qu_{m-1}+R, \nonumber  \\
         \rho^{(2)}&=&Su_{m-1}^3+Tu_{m-1}^2+Vu_{m-1}+U,  \nonumber \\
         \rho^{(3)}&=&Wu_m^2+Yu_m+Z,  \nonumber\eea
         with $|Q|=|R|=|T|=|V|=|U|=m-2,$ \ $|Y|=|Z|=m-1,$ \ and

\bea
P&=&a''+\frac{m^2-1}{24}\frac{(a')^2}{a}+\frac{(1-m)}{m}\frac{C}{a^m}a'
\nonumber\\
&-&\frac{1}{m}\frac{C'}{a^m}a+\frac{2(m-1)}{m^2}\frac{C^2}{a^{2m}}a
\label{sari} \\ \nonumber \\
S&=& a''a'-\frac{1}{6}\frac{(5+m)}{m}\frac{C}{a^m}a''a\nonumber
\\
&+&\frac{1}{6}\frac{(m-13)}{m}\frac{C'}{a^m}a'a+2\frac{(m-1)}{m^2}\frac{C^2}{a^{2m}}a'a
\nonumber \\
&+&
\frac{1}{6}\frac{(6m-m^2-5)}{m}\frac{C}{a^m}(a')^2+\frac{2}{m^2}\frac{C'C}{a^{2m}}a^2\nonumber
\\
&+& \frac{8}{3}\frac{(1-m)}{m^3}\frac{C^3}{a^{3m}}a^2  \label{lacivert}\\
\nonumber
\\
W&=& a(a')^2-120\frac{m+3}{m(m^3+3m^2-121m+597)}\frac{C}{a^m}a'a^2
\nonumber \\
&+&240\frac{m-1}{m^2(m^3+3m^2-121m+597)}\frac{C^2}{a^{2m}}a^3
\label{mavi} \eea
where \ $'=\frac{\partial{}}{\partial{u_{m-2}}}.$ Note that as
higher order conserved densities should be quadratic (Corollary
3.4) and top coefficients of conserved densities of consecutive
orders are related (Remark 3.5), we have
\be S=0 \quad \quad {\rm and}\quad \quad   W=a^2 P.\ee We will now
prove that, Proposition 4.4.1 together with the conservation
requirements of the $\rho^{(i)}$'s above will imply $A_{m-1}=C=0$.

\vskip 0.2cm

\noindent \textbf{Corollary 4.4.2} {\it Let \
$u_t=Au_m+Cu_{m-1}^2+Du_{m-1}+E,$ \ $m\geq 19,$ $m=2k+1$ \ with \
$A=a^m,$ \  $|A|=|C|=|D|=|E|=m-2.$ Then if $\rho^{(1)},$ \
$\rho^{(2)},$ \ and \ $\rho^{(3)}$ are conserved quantities, then}
\be A_{m-2}=C=0. \ee

\vskip 0.1cm

\noindent  {\it Proof.} Let's first consider  the case where
$C=0$. In this case, $P$ reduces to $P=(a')^2/a$, and substituting
this in $(\ref{fourtw})$, from $APa'/a=0$ we get $a'=0$, hence
$A_{m-2}=0$.

 If $C\ne 0$, from $(\ref{fourtw})$, $CP=0$ implies $P=0$. But as
$W=a^2P$, we have $W=0$ also.  Now, $(\ref{sari},\ref{lacivert},
\ref{mavi})$, is a system of three nonlinear differential
equations for $a$ and $C$.  We can in principle solve $a'$ from
$W=0$ in terms of $C$ as the root of a quadratic equation, then
substitute in $P=0$ and $S=0$ to get a system for $C$ and $C'$.
With symbolic computations it is possible to see that $C$ is zero.
We give below an analytic proof of this fact.

We define \ $\displaystyle \tilde{a}=a'/a$ \ and \ $\displaystyle
\tilde{C}=C/a^m$.  Then \ $ \displaystyle
C'/a^m=\tilde{C}'+m\tilde{a}\tilde{C}$ \ and \ $ \displaystyle
a''/a=\tilde{a}'+\tilde{a}^2$. Hence
\bea
\frac{P}{a}&=&\frac{m^2+23}{24}\tilde{a}^2+\frac{1-2m}{m}\tilde{a}\tilde{C}+\tilde{a}'\nonumber
\\
&+&\frac{2(m-1)}{m^2}\tilde{C}^2-\frac{1}{m}\tilde{C}', \\ \nonumber \\
\frac{S}{a^2}&=&
\tilde{a}^3+\frac{1}{6}\frac{(5m-m^2-10)}{m}\tilde{a}^2\tilde{C}\nonumber
\\
&+&
\frac{4m-2}{m^2}\tilde{a}\tilde{C}^2+\frac{1}{6}\frac{(m-13)}{m}\tilde{a}\tilde{C}\nonumber
\\
&-&
\frac{1}{6}\frac{(5+m)}{m}\tilde{a}'\tilde{C}+\tilde{a}\tilde{a}'+\frac{2}{m^2}
\tilde{C} \tilde{C}'\nonumber
\\
&+& \frac{8}{3}\frac{(1-m)}{m^3}\tilde{C}^3,\\
\nonumber \\
\frac{W}{a^3}&=&
\tilde{a}^2-120\frac{m+3}{m(m^3+3m^2-121m+597)}\tilde{a}\tilde{C}\nonumber
\\
&+&240\frac{m-1}{m^2(m^3+3m^2-121m+597}\tilde{C}^2. \label{fou34}
\eea
Note that $W/a^3$ is of the form  \bea
\frac{W}{a^3}=\tilde{a}^2-2\kappa_1
\tilde{C}\tilde{a}+\kappa_2\tilde{C}^2 \eea where $\kappa_1$ and
$\kappa_2$ are constants that can be identified from
$(\ref{fou34})$ and since \ $W=0$ \ \be
\tilde{a}_{1,2}=\tilde{C}(-\kappa_1\pm
\sqrt{\kappa_1^2-\kappa_2}). \label{turuncu} \ee If $\gamma $
denotes the coefficient of $C'$ in the expression above, we have
$\tilde{a}=\gamma \tilde{C}$ and $\tilde{a}'=\gamma \tilde{C}'$
and we obtain \bea
\displaystyle\frac{P}{a}&=&\left(\frac{m^2-23}{24}\gamma^2+\frac{1-2m}{m}\gamma+2\frac{m-1}{m^2}\right)\tilde{C}^2\nonumber
\\
&+& \left(\gamma -\frac{1}{m}\right)\tilde{C}'=0 \label{cay},\eea
\bea \displaystyle \tilde{C}^{-1}\frac{S}{a^2}&=&
\left(\gamma^3-\frac{\gamma^2}{3} \ \frac{4m+5}{m}+\gamma \
\frac{4m-2}{m^2}+\frac{8}{3} \ \frac{1-m}{m^3}\right)\tilde{C}^2
 \nonumber \\
&+&
\left(\gamma^2-\frac{3}{m}\gamma+\frac{2}{m^2}\right)\tilde{C}'=0.
\label{kahve} \eea
Now $(\ref{cay})$ and $(\ref{kahve})$ form a system of homogeneous
linear system for $\tilde{C}'$ and $C^2$. It can be checked that
the coefficient matrix is nonsingular, hence $\tilde{C}=0$. From
$(\ref{turuncu})$, we have $\tilde{a}=a'=0$, and the proof is
complete. \hfill  $\Box$

\subsection*{4.5 Step 5. Polinomiality in $u_{m-2}$, second result: $D_{m-2,m-2}=0$}

We will now prove that $D$ is linear in $u_{m-2}$

\noindent \textbf{Proposition 4.5.1} {\it Let
\be u_t=Au_m+Du_{m-1}+E, \ m \geq 19 \label{prop451},\ee
with $|A|=m-3, \ |D|=|E|=m-2,$ and assume that it admits a
conserved density of the form
\be \rho=Pu_{m-1}^2+Qu_{m-1}+R,  \label{prop452},\ee where
$|P|=m-3,$ $|Q|=|R|=m-2.$ Then} \be D_{m-2,m-2}=0. \ee \vskip
0.1cm
 \noindent  {\it Proof.} We compute \ $(\ref{bir})$
for equations $(\ref{prop451})$ and $(\ref{prop452})$ with $l=-1,$
$F_m=A,$ \ $\rho_{n,n}=2P$ to get: \bea
\left[k+\frac{1}{2}\right]2P_{m-3}Au_{m-2}-\left[k-\frac{1}{2}\right]2PA_{m-3}u_{m-2}=2PD
\label{yirmi} \eea
Differentiating \ $(\ref{yirmi})$ \ twice with respect to \
$u_{m-2}$ we obtain
\be PD_{m-2,m-2}=0 \ee \hfill  $\Box$
\vskip 0.1cm
It can be checked that the canonical density
$\rho^{(3)}$ in $(A.5)$ is of the form $(\ref{prop452})$ and the
coefficient $P$ of the quadratic top derivative $u_{m-1}^2,$  is
as follows:

\bea P&=&\frac{m^4-10m^2+720m-2151}{m^4-10m^2+9}a a_{m-3}^2
\nonumber \\
&+& \frac{60(-3m^2+8m+3)}{m(m^4-10m^2+9)a^m}a^2a_{m-3}D_{m-2}
\nonumber \\
&+& \frac{60}{m^2(m^2+4m+3)a^{2m}}a^3D_{m-2}^2 \label{siyah}\eea

That is \be P=P^{(0)}+P^{(1)}D'+P^{(2)}D'^2 \ee Hence $PD''=0$
implies \be D''(P^{(0)}+P^{(1)}D'+P^{(2)}D'^2)=0\ee From this it
follows that \be D''=D_{m-2,m-2}=0 \ee

\vskip 0.2cm

\noindent \textbf{Corollary 4.5.2} {\it If the canonical density
$\rho^{(3)}$ is conserved $D$ is linear in $u_{m-2}$.}

\vskip 0.2cm
Thus we conclude that an integrable evolution equation should be
of the form \be u_t=Au_m+Gu_{m-1}u_{m-2}+H u_{m-1}+E, \ee where
$|A|=|G|=|H|=m-3$ and $|E|=m-2.$

\subsection*{4.6 Step 6. Polinomiality in $u_{m-2}$, third result:
$E_{m-2,m-2,m-2,m-2}=0$. }

The final step is to  determine the $u_{m-2}$ dependency in $E$
using the explicit form of $\rho^{(1)}$.  For  \be u_t=Au_m+G
u_{m-1}u_{m-2}+H u_{m-1}+E, \ m \geq 19, \label{prop6} \ee with
$|A|=|G|=|H|=m-3$ and $|E|=m-2$, the canonical density
$\rho^{(1)}$ is of the form \bea \rho^{(1)}&=&
\frac{u_{m-2}^2}{m^2-1}
\left[\frac{a_{m-3}^2}{a}(m^2-1)+36\frac{a_{m-3}G}{a^m}(\frac{1}{m}-1)+24\frac{aG_{m-3}}{ma^m}+12\frac{aG^2}{ma^{2m}}(1-\frac{1}{m})\right]
\nonumber \\
&+&
\frac{2u_{m-2}u_{m-3}}{m^2-1}\left[\frac{a_{m-3}a_{m-4}}{a}(m^2-1)+18\frac{a_{m-4}G}{a^m}(\frac{1}{m}-1)+12\frac{G_{m-4}a}{ma^m}\right]
\nonumber \\
&+&
12u_{m-2}\frac{H}{a^{2m}m^2(m+1)}\left[2aG-ma^ma_{m-3}\right]\nonumber
\\
&+&
u_{m-3}^2\frac{a_{m-4}^2}{a}-12u_{m-3}\frac{a_{m-4}H}{m(m+1)a^m}\nonumber
\\
&+&
12\frac{a}{m^2(m^2-1)a^{2m}}\left[-2ma^mE_{m-2}+H^2(m-1)\right]
\label{prop66} \eea where $a^m=A$

\noindent \textbf{Proposition 4.6.1} {\it Let \ \be u_t=Au_m+G
u_{m-1}u_{m-2}+H u_{m-1}+E, \ m \geq 19, \label{prop6} \ee with
$|A|=|G|=|H|=m-3$ and $|E|=m-2$,  Then
 if $\rho^{(1)}$ is a conserved quantity, then}
 \be  E_{m-2,m-2,m-2,m-2}=0. \ee
\vskip 0.1cm

\noindent {\it Proof.} Notice that $\rho^{(1)}$ has order $m-2$
but it is not a priori even polynomial in \ $u_{m-2}$ \ since we
don't know the form of \ $E.$ We substitute $(\ref{prop6})$ in
$(\ref{bir})$ with $l=-2,$ \ $F_m=A$ \ and \

 $F_{m-1}=Gu_{m-2}+H$
\be (k+\frac{1}{2})AD\rho^{(1)}_{m-2,m-2}-(k-\frac{3}{2})DA
\rho^{(1)}_{m-2,m-2}=(Gu_{m-2}+H)\rho^{(1)}_{m-2,m-2}
\label{prop666} \ee The coefficient of $u_{m-1}$ in
$(\ref{prop666})$ is: \be \rho^{(1)}_{m-2,m-2,m-2}=0, \ee
hence since $a \neq 0$, it follows that
\be
\frac{\partial^3{\rho^{(1)}}}{\partial{u_{m-2}}^3}=-\frac{24a^{-m+1}}{m(m^2-1)}E_{m-2,m-2,m-2,m-2}=0.
\ee \hfill  $\Box$

\vskip0.05cm
\noindent
\textbf{Result C: Polynomiality in $u_m, u_{m-1}, u_{m-2}$}
\vskip 0.05cm \noindent \textbf{Corollary 4.6.2} {\it
If the canonical densities  $\rho^{(1)}$, $\rho^{(2)}$  and $\rho^{(3)}$ are conserved quantities
for the evolution equation
$u_t=F$,
then
\be
F=Au_m+Gu_{m-2}u_{m-1}+Hu_{m-1}+Ju_{m-2}^3+Lu_{m-2}^2+Nu_{m-2}+S
\label{conclusion}\ee
where $A$, $G$, $H$, $J$, $L$, $N$, $S$   are functions of $x,t,u,\dots u_{m-3}$.}
\vskip 0.2cm

\noindent \textbf{Remark 4.7} By the use of the conserved
densities $\rho^{(i)}$,  \ $i=1,2,3$ we have shown that $A$ is
independent of $u_{m-3}$ and $G=0$. However it was not possible to
obtain polynomiality in $u_{m-3}$ because further conserved
densities were needed.

\section*{5. CONCLUSION}

In the present paper we obtained all the polynomiality information
that could be extracted from the conserved densities up to
$\rho^{(3)}$ (See Remark 4.7). In general computations we haven't
used $\rho^{(-1)}$ and $\rho^{(0)}$, but using these for $m=7$
didn't give any further information.

For $m=7$, in initial computations we used dependencies in all
derivatives. But at later stages this was impossible and in the
search of computationally efficient methods, we noticed that all
polynomiality results followed from the coefficient of the top
order term in $(\ref{bir})$ and $(\ref {iki})$.
 The use of only the top dependency lead us to formulate a graded
algebra structure on the polynomials in the derivatives $u_{k+i},$
that we called ``level grading" \cite{tez}. This structure is
based on the fact that derivatives of a function of
$x,t,u,\dots,u_k$ are polynomial in higher order derivatives and
have a natural scaling by the order of differentiation above the
``base level $k.$" The crucial point is that equations relevant
for obtaining polynomiality results involve only the terms with
top scaling weight with respect to level grading. Thus one can
work with the dependency on the top level term only and reduce the
scope of symbolic computations to a feasible range. Applications
of the ``level grading'' structure to the classification problem
will be discussed elsewhere.

By the remark following Proposition 3.1, the polynomiality results
obtained in Section 4 are valid for $m\ge 19$. For
$m=7,9,11,13,15,17$ the conserved density conditions are computed,
directly, {\it without using} $(\ref{bir})$ and $(\ref {iki})$,
with the symbolic programming language REDUCE and it is shown that
the evolution equation have again the form given by
(\ref{conclusion}).  The results of Corollary 4.6.2 are valid for
all $m\ge 7$ and we restate it here for convenience.

\vskip 0.2cm \noindent  \textbf{Corollary 4.6.2} {\it Let
$$u_t=F(x,t,u,\dots ,u_m)$$
be a scalar  evolution equation in one space dimension of order $m=2k+1$ where $m\ge 7$.
If the canonical densities  $\rho^{(i)}$, $i=1,2,3$ given in the Appendix are conserved quantities then
$$F=Au_m+Gu_{m-2}u_{m-1}+Hu_{m-1}+Ju_{m-2}^3+Lu_{m-2}^2+Nu_{m-2}+S.$$
where $A$, $G$, $H$, $J$, $L$, $N$, $S$   are functions of $x,t,u,\dots u_{m-3}$.}

\vskip 0.2cm

For future work, the graded algebra structure seems to be
promising tool both for the computation of conserved densities and
for obtaining integrability conditions. In fact, conserved densities for seventh
order equations have been computed  and it has been observed that
there are candidates for integrable equations where $A$ may depend on $u_4$ and is
non-polynomial in $u_4$. In this sense,
Corollary 4.6.2 is the best polynomiality result for evolution equations of order 7.

\vskip 1cm

\appendix.

\noindent {\bf APPENDIX  A: Canonical Densities of the Formal
Symmetry Method}
\vskip 0.2cm
\noindent If the evolution equation $u_t=F[u]$ is integrable, it
is known that the quantities
$$\rho^{(-1)}=F_m^{-1/m}, \quad  \rho^{(0)}=F_{m-1}/F_m,
\eqno(A.1a) $$
where
$$F_m=\frac{\partial{ F}}{\partial{u_m}}, \quad
F_{m-1}=\frac{\partial{F}}{\partial{u_{m-1}}}
\eqno (A.1b)$$
are conserved densities for equations of any order \cite{MSS91}.
The next three canonical densities {\it up to total derivatives}
computed in \cite{B2005} are presented below. The expressions of
the canonical densities up to total derivatives are convenient for
the present work   but one has to be careful in general in adding
a total derivative to the canonical densities. Because if
$\rho^{(i)}$ is a canonical density with
$D_t\rho^{(i)}=D\sigma^{(i)}$, then $\rho^{(i+k)}$ will involve
$\sigma^{(i)}$ after some $k$ \cite{MSS91}. Hence although total
derivatives in $\rho^{(i)}$ are irrelevant for the existence of
$\sigma^{(i)}$, they should not be omitted whenever $\sigma^{(i)}$
enters in the expression of another canonical density. We shall
use the following notation
$$
a=F_m^{1/m}, \quad \alpha_{(i)}=\frac{F_{m-i}}{F_m}, \ \ i=1,2,3,4.
\eqno(A.2)$$
\bea
\rho^{(1)}&=&a^{-1}(Da)^2-\frac{12}{m(m+1)}Da\alpha_{(1)}+a\left[\frac{12}{m^2(m+1)}\alpha_{(1)}^2-\frac{24}{m(m^2-1)}\alpha_{(2)}\right],
\label{roro1} \quad \quad \quad \quad (A.3) \nonumber\eea
\bea
\rho^{(2)}&=&a(Da)\left[D\alpha_{(1)}+\frac{3}{m}\alpha_{(1)}^2-\frac{6}{(m-1)}\alpha_{(2)}\right]
\nonumber \\
&+&2a^2\left[-\frac{1}{m^2}\alpha_{(1)}^3+\frac{3}{m(m-1)}\alpha_{(1)}\alpha_{(2)}-\frac{3}{(m-1)(m-2)}\alpha_{(3)}\right],
\label{roro2} \quad \quad \quad \quad \quad \quad \quad \quad
\quad (A.4)\nonumber \eea
\bea
\rho^{(3)}&=&a(D^2a)^2-\frac{60}{m(m+1)(m+3)}a^2D^2aD\alpha_{(1)}+\frac{1}{4}a^{-1}(Da)^4
\nonumber \\
&+&30a(Da)^2\left[\frac{(m-1)}{m(m+1)(m+3)}D\alpha_{(1)}+\frac{1}{m^2(m+1)}\alpha_{(1)}^2-\frac{2}{m(m^2-1)}\alpha_{(2)}\right]\nonumber
\\
&+&\frac{120}{m(m^2-1)(m+3)}a^2Da\left[-\frac{(m-1)(m-3)}{m}\alpha_{(1)}D\alpha_{(1)}+(m-3)D\alpha_{(2)}\right.
\nonumber \\
&-&
\left.\frac{(m-1)(2m-3)}{m^2}\alpha_{(1)}^3+\frac{6(m-2)}{m}\alpha_{(1)}\alpha_{(2)}-6\alpha_{(3)}\right]
\nonumber \\
&+&\frac{60}{m(m^2-1)(m+3)}a^3\left[\frac{(m-1)}{m}(D\alpha_{(1)})^2-\frac{4}{m}D\alpha_{(1)}\alpha_{(2)}+\frac{(m-1)(2m-3)}{m^3}\alpha_{(1)}^4\right.
\nonumber \\
&-&
\left.4\frac{(2m-3)}{m^2}\alpha_{(1)}^2\alpha_{(2)}+\frac{8}{m}\alpha_{(1)}\alpha_{(3)}+\frac{4}{m}\alpha_{(2)}^2-\frac{8}{(m-3)}\alpha_{(4)}\right].
\label{roro3} \quad \quad \quad \quad \quad \quad \quad \quad
 \quad \quad (A.5) \nonumber \eea

\newpage
\centering
 \textbf{Evolution equation of order $m$ with generic and
 canonical densities}
\begin{tabular}{|l|l|l|l|l|}
  \hline
  Step & Evolution Equation & Type of the       & Form of the       & Result \\
       &                    & conserved density & conserved density   &  \\

  \hline
      &            &         &                        &  \\
  $1$ & $u_t=F[u]$ & generic & $\rho=Pu_{m+1}^2+Qu_{m+1}+R$ & $\frac{\partial^2{F}}{\partial{u_m}^2}=0$ \\
      &            &         &                              &                                           \\
  $2$ & $u_t=Au_m+B$ & generic & $\rho=Pu_m^2+Qu_m+R$ & $\frac{\partial{A}}{\partial{u_{m-1}}}=0$ \\
      &              &         &                      &                                           \\
  $3$ & $u_t=Au_m+B$ & canonical & $\rho=\rho^{(3)}$ & $\frac{\partial^3{B}}{\partial{u_{m-1}}^3}=0$ \\
      &              &           &                   &                                               \\
  $4$ & $u_t=Au_m+Cu_{m-1}^2$ & generic  and & $\rho=Pu_{m-1}^2+Qu_{m-1}+R$& $\frac{\partial{A}}{\partial{u_{m-2}}}=0,$\\
%      &                       & and          &           & \\
      &  $\ \ \ +Du_{m-1}+E$  & canonical         &  $\eta=Su_m^2+Tu_m+U$                 &  $C=0$  \\
      &                       &          &                &   \\
  $5$ & $u_t=Au_m+Du_{m-1}$ & generic and & $\rho=Pu_{m-1}^2+Qu_{m-1}+R$ & $\frac{\partial^2{D}}{\partial{u_{m-2}}^2}=0$ \\
%      &                  &            &               &     \\
      &  $\ \ \ +E$                      & canonical        &          &            \\
      &           &           &           &   \\
  $6$ & $u_t=Au_m$ & canonical & $\rho=\rho^{(1)}$ & $\frac{\partial^4{E}}{\partial{u_{m-2}^4}}=0$ \\
 %     &          &              &                 &      \\
      &  $ \ \ \ \ +Gu_{m-1}u_{m-2}$                          &           &     &     \\
%      &           &             &                 &         \\
      & $ \ \ \ \ +Hu_{m-1}+E$     &    &                   &    \\
  \hline
\end{tabular}
 Table1:Polynomiality results for the general case

\newpage
\appendix

\noindent {\bf APPENDIX B: Expressions of $k$'th Order
Derivatives}\cite{B2005}

\bea D^k\varphi &=& \varphi_nu_{n+k}+P(u_{n+k-1}), \ \ k\geq 1
\quad \quad \quad \quad \quad \quad \quad \quad \quad \quad \quad \quad \quad \quad \quad \quad \quad \quad \quad \quad \quad \quad (A.6a) \nonumber \\ \nonumber \\
\displaystyle D^k\varphi &= & \varphi_n
u_{n+k}+\left[\varphi_{n-1}+kD\varphi_n\right]u_{n+k-1}+P(u_{n+k-2}),
\
\ k\geq 3 \quad \quad \quad \quad \quad \quad \quad \quad \quad \quad \quad (A.6b) \nonumber \\ \nonumber \\
\displaystyle D^k\varphi &=&
\varphi_nu_{n+k}+\left[\varphi_{n-1}+kD\varphi_n
\right]u_{n+k-1} \nonumber \\
&+& \displaystyle \left[\varphi_{n-2}+kD\varphi_{n-1}+{k \choose
2}D^2\varphi_n\right]u_{n+k-2}+P(u_{n+k-3}), \ \ k\geq 5 \quad
\quad \quad \quad \quad \quad \quad \quad \quad (A.6c) \nonumber
\\ \nonumber \\
\displaystyle D^k\varphi &=&
\varphi_nu_{n+k}+\left[\varphi_{n-1}+kD\varphi_n\right]u_{n+k-1}
\nonumber \\
&+& \left[\varphi_{n-2}+kD\varphi_{n-1}+{k \choose
2}D^2\varphi_n\right]u_{n+k-2} \nonumber \\
&+& \left[\varphi_{n-3}+kD\varphi_{n-2}+{k \choose
2}D^2\varphi_{n-1}+{k \choose 3}D^3\varphi_n\right]u_{n+k-3}
\nonumber \\
&+& P(u_{n+k-4}), \ \ k \geq 7 \quad \quad \quad \quad \quad \quad
\quad \quad \quad \quad \quad \quad \quad \quad \quad \quad \quad
\quad \quad \quad \quad \quad \quad \quad \quad \quad \quad (A.6d)
\nonumber \eea

\newpage


\begin{thebibliography}
{99}

\bibitem[1]{SW98} J.A. SANDERS and J.P. WANG, On the integrability of homogeneous
scalar evolution equations, {\it Journal of Differential
Equations} 147:410-434 (1998).

\bibitem[2]{SW2000} J.A. SANDERS and J.P. WANG, On the
integrability of non-polynomial scalar evolution equations, {\it
Journal of Differential Equations}, 166:132-150 (2000).

\bibitem[3]{MSS91} A.V. MIKHAILOV, A.B. SHABAT and V.V SOKOLOV.
The symmetry approach to the classification of integrable
equations in {\it What is Integrability?} edited by V.E. Zakharov,
Springer-Verlag, Berlin, 1991.

\bibitem[4]{HSS95} R.H. HEREDERO, V.V. SOKOLOV and S.I. SVINOLUPOV,
Classification of 3rd order integrable evolution equations, {\it
Physica D}, 87:32-36 (1995).

\bibitem[5]{B2005}  A.H.BILGE, Towards the Classification of Scalar
Non-Polinomial Evolution Equations: quasi-linearity, {\it Computers
and Mathematics with Applications}, 49:1837-1848, (2005).

\bibitem[6]{O93} P.J. OLVER, Evolution equations possessing
infinitely many symmetries, {\it Journal of Mathematical Physics},
18:1212-1215, (1977)

\bibitem[7]{A79} M. ADLER, On a trace functional for formal
pseudo-differential operators and the symplectic structure of the
Korteweg-de Vries type equations, {\it Invent. Math.}, 50:219-248,
(1979)

\bibitem[8]{tez}E. MIZRAHI, {\it Towards the Classification of Scalar Integrable
Evolution Equations in $(1+1)$ Dimensions}, Ph.D. Thesis, Istanbul
Technical University, June 2008.
\end{thebibliography}
\end{document}